\documentclass{article}
\usepackage{spconf}
\usepackage{cite}
\usepackage{amsmath,amssymb,amsfonts}
\usepackage{algorithmic}
\usepackage{graphicx}
\usepackage{textcomp}
\usepackage[dvipsnames]{xcolor}
\PassOptionsToPackage{hyphens}{url}
\usepackage{colortbl}
\usepackage{enumitem}
\usepackage{subfigure}
\usepackage[unicode,plainpages=false,pdfpagelabels,breaklinks,colorlinks]{hyperref}
\usepackage{soul}
\usepackage{makecell}
\hypersetup{
  linkcolor=BrickRed,
  citecolor=Green,
  filecolor=Mulberry,
  urlcolor=NavyBlue,
  menucolor=BrickRed,
  runcolor=Mulberry,
  linkbordercolor=BrickRed,
  citebordercolor=Green,
  filebordercolor=Mulberry,
  urlbordercolor=NavyBlue,
  menubordercolor=BrickRed,
  runbordercolor=Mulberry
}

\usepackage{xurl}

\title{TTM-Bench: A Framework for Text-to-Music System Performance Benchmarking 
}

\name{
Giorgia Adorni$^{1}$ \qquad
Michela Papandrea$^{1}$ \qquad
Battista Rimoldi$^{2}$ \qquad
Tiziano Leidi$^{1}$
}

\address{
$^{1}$Institute of Information Systems and Networking (ISIN), SUPSI, Lugano, Switzerland\\
$^{2}$t2b AG, Basel, Switzerland
}

\begin{document}

\maketitle

\begin{abstract}
Text-to-music (TTM) systems are increasingly used to generate musical audio from natural-language descriptions.
Robust evaluation is therefore essential, yet reliable performance comparison remains challenging. This difficulty stems from differences in system architecture, supported conditioning information, and access mode, as well as heterogeneous and fragmented metrics that cannot be applied uniformly across systems.
To address these challenges, we introduce \textbf{TTM-Bench}, a framework that defines a common protocol for systematic, reproducible performance benchmarking of contemporary TTM systems. 
It evaluates performance along two dimensions: \textit{musical-content alignment}, quantified by interpretable semantic, genre, and musical-descriptor agreement scores against a common musical specification and summarized by an aggregate score; and \textit{computational efficiency}, characterized by generation latency and real-time factor, alongside resource use for local models and cost for hosted services. 
We demonstrate the framework through a preliminary comparative case study, illustrating the complementary evidence captured by these dimensions.
The results show that higher musical-content alignment does not systematically coincide with lower computational demands, highlighting the importance of assessing TTM performance through distinct, interpretable measures rather than a reductive overall indicator.
\end{abstract}

\begin{keywords}
text-to-music generation, benchmarking framework, musical-content alignment, computational efficiency, automatic evaluation

\end{keywords}

\section{Introduction}
\label{sec:introduction}
Text-to-music (TTM) generation has emerged as a prominent research direction in generative audio, allowing systems to synthesize music from textual descriptions and other music-related conditioning information.
Recent advances in this field have led to TTM systems with substantially different conditioning formats, ranging from free-form prompts to structured musical attributes.
This heterogeneity poses a comparability problem: the same prompt may not express the intended musical content equivalently across systems, whereas system-specific reformulation may alter the underlying specification.
A benchmarking protocol must therefore preserve a shared musical specification while expressing it in a form compatible with each system.
Beyond conditioning, TTM systems also differ in architecture, supported output duration, and execution mode. These differences affect how computational efficiency can be measured and compared: local systems allow direct measurement of hardware-level resource use, whereas hosted APIs allow only service-level observations such as latency and cost.
%
A further source of heterogeneity concerns evaluation. Existing automatic metrics target different performance dimensions and therefore capture separate aspects of system behavior. Contrastive Language-Audio Pretraining (CLAP)-based similarity measures semantic correspondence between a text prompt and generated audio~\cite{wu2023clap}, whereas Fréchet Audio Distance (FAD) compares generated and reference audio distributions and therefore depends on the choice of reference corpus~\cite{kilgour2019fad}. As a result, metric selection depends on the benchmarking objective and affects how system performance is characterized.
%
To the best of our knowledge, available benchmarks address specific aspects of TTM evaluation under relatively controlled settings. However, they do not directly support systematic comparison of existing systems that differ in conditioning formats, architectures, and access modes.

In this work, we make two main contributions. 
First, we introduce \textbf{TTM-Bench}, a framework for benchmarking heterogeneous TTM systems across two distinct dimensions. 
It establishes a common protocol that preserves the underlying musical specification while adapting its representation to each system and explicitly separates musical-content alignment from access-aware computational efficiency.
TTM-Bench integrates complementary evidence within two workflows.
The \textit{musical-content alignment} workflow evaluates how closely each generated track matches the intended musical content, using semantic correspondence with the conditioning text and genre and musical-descriptor similarity to the reference audio. It retains interpretable component scores alongside an aggregate alignment score.
The \textit{computational efficiency} workflow characterizes generation time, resource use, and cost within the observable boundaries of local and hosted execution.
Second, we illustrate the framework through a \textit{preliminary case study} involving 13 open and commercial TTM systems.

\section{Related Work}
\label{sec:prior-work}
Research on TTM performance benchmarking spans four areas: automatic metrics, learned perceptual evaluators, computational efficiency studies, and cross-system benchmarks.

{Automatic metrics} quantify properties such as semantic correspondence with the conditioning text, differences between generated and reference audio distributions in learned embedding spaces, and audio-derived musical characteristics~\cite{wu2023clap,kilgour2019fad,yang2020evaluation,lerch2025survey}. These measures capture different aspects of generated audio, and their results depend on methodological choices such as the embedding model, reference data, preprocessing, sample size, and aggregation~\cite{gui2024fad}. 

{Learned perceptual evaluators}, such as MusicEval, MuQ-Eval, SongBench, AudioEval, and TuneJury, use human-annotated data to estimate perceived text-audio alignment, listener preference, and musical quality across aspects including vocals, composition, arrangement, and production~\cite{liu2025musiceval,zhu2026muqeval,wu2026songbench,wang2026audioeval,kim2026tunejury}.
These approaches complement automatic metrics by capturing perceptual aspects that are difficult to measure directly, but their reliability depends on the quality, coverage, and potential biases of the human judgments used for training.

{Computational efficiency studies} mainly characterize generation speed through measures such as inference latency and real-time factor~\cite{evans2024stableaudio,ning2025diffrhythm,gong2025acestep}. These results are often reported under different hardware and experimental settings, limiting direct comparison, while other relevant factors such as memory use, energy consumption, and hosted-service cost are less consistently considered.

{Cross-system benchmarks} address different sources of evaluation variability.
Gr{\"o}tschla et al. relate model and metric outputs to human preferences, Music Arena facilitates live pairwise preference evaluation with LLM-based routing across systems with different conditioning formats, and the ATTM Grand Challenge standardizes training data and constraints in a fair-play setting ~\cite{grotschla2025benchmarking,kim2025musicarena,hsieh2026attm}. 
These approaches emphasize preference-based or controlled evaluation, leaving already-available systems with heterogeneous conditioning and access modes less systematically addressed.

TTM-Bench instead targets already-available systems with heterogeneous conditioning formats and access modes by evaluating musical-content alignment at the component level and computational efficiency using the measurements observable under each access mode.


\section{Method}
\label{sec:method}
The benchmarking procedure is organized into two workflows, described below.
We will make the code, descriptor manifests, system-specific prompts, derived metrics, and analysis notebooks publicly available on Zenodo.

\begin{figure}[htb]
  \centering
  \includegraphics[width=\linewidth]{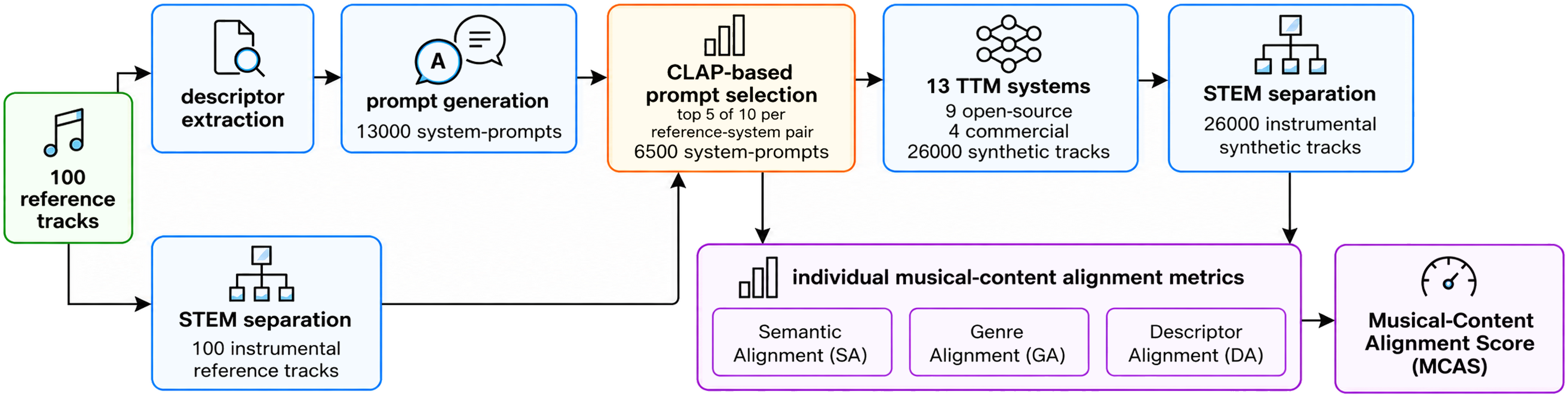}
  \caption{Musical-content alignment benchmarking workflow.}
  \label{fig:musical-content-alignment-workflow}
\end{figure}
The \textbf{musical-content alignment workflow} (Fig.~\ref{fig:musical-content-alignment-workflow}) assesses how well generated tracks match a common reference-derived musical specification, expressed in the conditioning format supported by each system. 
We characterize alignment along three dimensions.
(1) \textit{Semantic Alignment} (SA) measures semantic similarity between the conditioning text and generated audio using LAION-CLAP cosine similarity~\cite{wu2023clap}.
(2) \textit{Genre Alignment} (GA) combines linearly genre-set overlap (20\%) and classifier-score similarity (80\%). 
We estimate genre information using three pretrained classifiers with different taxonomies: Discogs519~\cite{alonso2023maest}, Discogs400~\cite{alonso2022discogs}, and MTG-Jamendo~\cite{bogdanov2019mtgjamendo}. The first two predict 519 and 400 styles from the Discogs taxonomy, respectively, while MTG-Jamendo uses a separate 87-genre taxonomy. Using multiple classifiers reduces dependence on a single genre representation.
We compute genre-set overlap as the Jaccard similarity between high-level genre sets predicted for the reference and generated audio by the two Discogs classifiers. We calculate classifier-score similarity by combining cosine similarities between the corresponding genre-score vectors from all three classifiers weighted at 50\%, 25\%, and 25\%, respectively.
(3) \textit{Descriptor Alignment} (DA) combines timbre-category (50\%), BPM (40\%), and tempo-label (10\%) similarity, based respectively on a fixed distance matrix over warm, bright, and harsh, relative BPM difference, and normalized ordinal distance from slow to fast.
These measures evaluate correspondence at the individual-generation level. We do not include distribution-level metrics such as FAD, as they compare audio sets rather than individual generations with their intended musical specification.
We calculate the \textit{Musical-Content Alignment Score} (MCAS) as a linear combination of the three components. It is not intended as an overall measure of TTM systems' performance.
\begin{equation}
\mathrm{MCAS}=0.40\cdot\mathrm{SA}+0.35\cdot\mathrm{GA}+0.25\cdot\mathrm{DA} .
\label{eq:mcas}
\end{equation}
We fix the weights empirically, assigning greater weight to SA for overall conditioning correspondence than to the narrower musical properties captured by GA and DA.
We assess MCAS weight sensitivity using equal weighting, local perturbations, and leave-one-component-out configurations, and measure agreement with the proposed weighting through Spearman's \(\rho\) and Kendall's \(\tau\).
We provide implementation details and normalization functions with the released code.


\begin{figure}[htb]
  \centering
  \includegraphics[width=\linewidth]{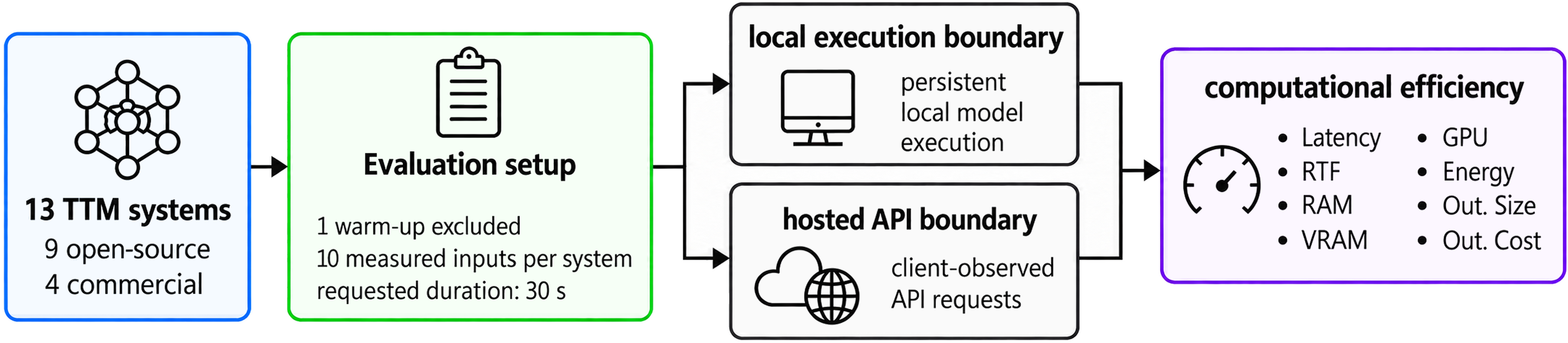}
  \caption{Computational efficiency benchmarking workflow.}
  \label{fig:computational-evaluation-workload}
\end{figure}
The \textbf{computational efficiency workflow} (Fig.~\ref{fig:computational-evaluation-workload}) characterizes generation time, resource use, and cost within the measurement boundaries of local execution and hosted services.
For local execution, latency covers model-side conditioning through construction of the usable waveform, with CUDA synchronization used for timing. 
For hosted services, it covers request submission through audio reception and decoding and may therefore include provider-side generation, queuing, network transfer, polling, download, and local postprocessing. 
When a system cannot natively generate the target duration, efficiency is measured for the computation required to obtain the standardized $30s$ output.
For both access types, real-time factor (RTF) is computed as measured latency divided by usable output duration. 
The reported measures reflect what is observable under each access mode. Local execution provides latency, RTF, memory use, GPU utilization, and GPU energy consumption, whereas hosted services provide client-observed latency, RTF, and provider-listed API cost. We do not estimate unavailable server-side resources.

\begin{figure}[htb]
  \centering
  \includegraphics[width=\linewidth]{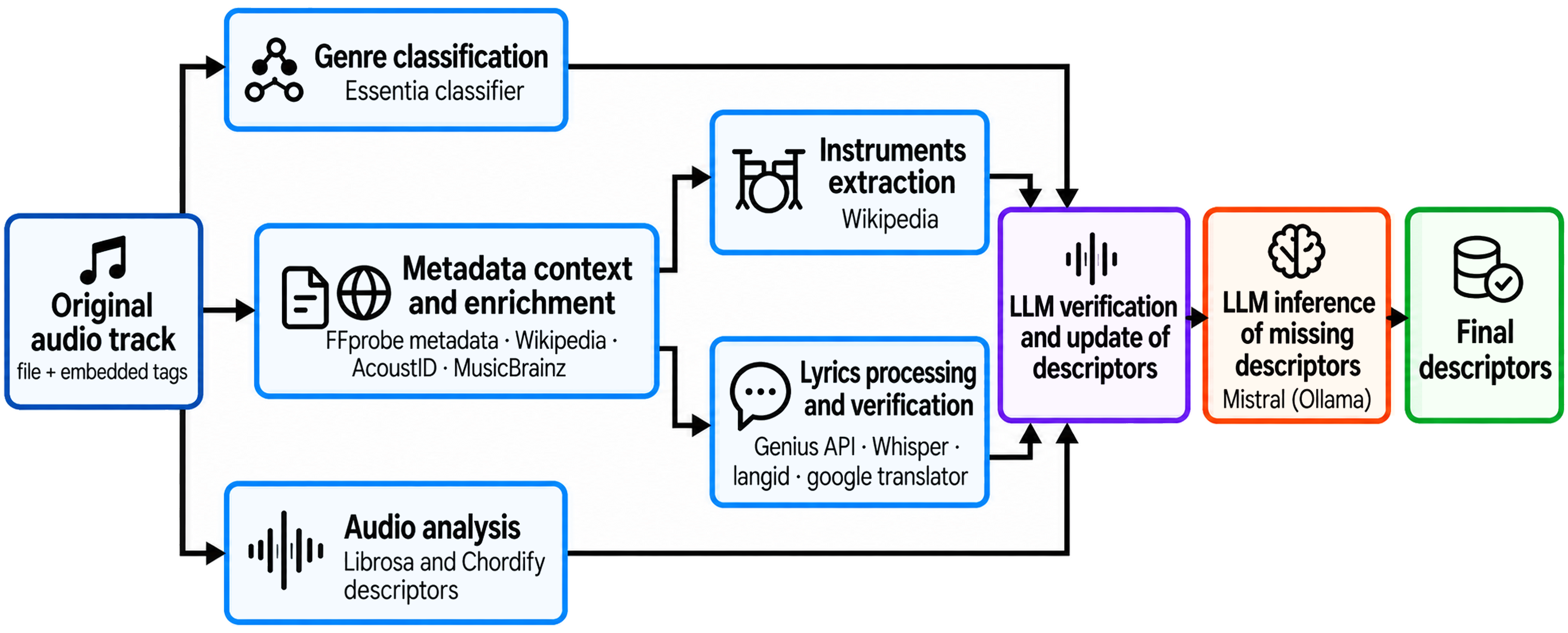}
  \caption{Descriptor-extraction pipeline.}
  \label{fig:descriptor-extractor}
\end{figure}

To validate the proposed TTM-Bench, we conduct a \textbf{preliminary case study} and analyze the results. 
We restrict the evaluation to text-conditioned instrumental generation by using HTDemucs to extract instrumental stems from both reference and generated audio, reducing the influence of differences in vocal or lyric generation on the comparison~\cite{rouard2023htdemucs}.
We evaluate 13 TTM systems spanning different conditioning formats, access modes, and architectural families, including autoregressive, diffusion, flow-matching, and hybrid approaches.
The set includes 9 open-source models: ACE-Step v1 3.5B~\cite{gong2025acestep}, AudioLDM2 Music~\cite{liu2024audioldm2}, DiffRhythm~\cite{ning2025diffrhythm}, HeartMuLa OSS 3B~\cite{yang2026heartmula}, InspireMusic 1.5B Long~\cite{zhang2025inspiremusic}, MusicGen Large~\cite{copet2023musicgen}, MusicLDM~\cite{chen2024musicldm}, Stable Audio Open 1.0~\cite{evans2024stableaudioopen}, and YuE~\cite{yuan2025yue}; and 4 commercial APIs: Lyria 2~\cite{google2025lyria2} and Lyria 3~\cite{google2026lyria3}, accessed via fal.ai, Stable Audio 2.5~\cite{stability2025stableaudio25}, and Stable Audio 3~\cite{evans2026stableaudio3}.
%
The reference collection comprises $100$ commercially released tracks obtained from legitimate music sources. We curated the collection to span multiple decades, genres, and tempo ranges, providing heterogeneous case-study conditions rather than statistical representativeness.
We release only the reference metadata and derived descriptors, not the original reference audio files.
Our extraction pipeline (Fig.~\ref{fig:descriptor-extractor}) derives 46 attributes per track from metadata, audio analysis, and controlled descriptor enrichment, including established music information retrieval (MIR) tools and models~\cite{bogdanov2013essentia,alonso2020tensorflow}.
For each reference-system pair (100 reference tracks $\times$ 13 TTM systems), we restrict the common musical specification to the attributes supported by that system. We use \texttt{Llama 3.1 8B}~\cite{grattafiori2024llama3} to generate 10 prompt variants in the required input format, ranging from comma-separated attribute lists to natural-language descriptions of varying detail.
We then validate each to ensure it preserves the selected attribute values without introducing unsupported musical information.
For each pair, we retain the 5 prompt variants with the highest LAION-CLAP similarity to the instrumental audio reference~\cite{wu2023clap}, yielding \(100\times13\times5=6500\) prompts. 
Finally, we generate 4 $30s$ tracks from each retained prompt, resulting in \(6500\times4=26000\) tracks. We compute SA, GA, DA, and MCAS for each, average the score first within each reference-system pair, and then across references to obtain system-level scores.
We assess robustness to reference-set composition through paired reference-level bootstrap resampling with $10000$ iterations. At each iteration, we sample $100$ references with replacement using the same reference indices for all systems, and recompute system-level scores and pairwise contrasts.
We assess the complementarity of SA, GA, and DA using Pearson and Spearman correlations at both generation and system levels. We interpret these correlations descriptively because outputs within reference-system pairs are not independent and only 13 systems are evaluated.
%
%
We evaluate computational efficiency after one unmeasured warm-up using 10 distinct inputs per system, targeting $30s$ outputs.
For local runs, we use an NVIDIA RTX PRO 6000 Blackwell Server Edition (96~GB); we document hardware, software, and system-specific inference configurations in the Zenodo benchmark manifest. 
We sample resource measurements at approximately $10 Hz$ during generation, excluding the time required to save the final audio.
We define peak RAM as the maximum memory used by the generation process and its child processes, and peak VRAM as the maximum GPU memory reported by \texttt{nvidia-smi}. 
We average GPU utilization across samples and estimate GPU-board energy by integrating the reported power draw over time.
We do not measure CPU or whole-system energy.

\section{Results}
\label{sec:results}
Table~\ref{tab:components} summarizes SA, GA, DA, and MCAS for the evaluated systems.
Across the tested alternatives, agreement with the proposed weighting remains high (median/minimum Spearman \(\rho=0.978/0.929\); Kendall \(\tau=0.923/0.821\)). This supports the adopted formulation as a reasonable choice, with relative system-level MCAS results remaining stable under the tested weighting alternatives.
Commercial systems outperform open models across all musical-content alignment measures. Within the former group, Stable Audio~2.5 leads in SA and GA, while Lyria~3 records the highest DA; within the latter, Stable Audio Open~1.0 leads in SA and DA, with MusicGen Large obtaining the highest GA. These differences indicate that relative performance varies across the individual alignment components.
For the three smallest pairwise MCAS contrasts, the 95\% bootstrap intervals for the pairwise difference between Stable Audio~2.5 and Lyria~3 (\([-0.012,0.013]\)), Lyria~3 and Stable Audio~3 (\([-0.008,0.014]\)), and MusicGen Large and Stable Audio Open~1.0 (\([-0.016,0.020]\)) all include zero. Thus, small differences in MCAS are not robust to changes in reference-set composition.
At generation level, SA, GA, and DA show weak positive pairwise correlations (Pearson \(r=0.217\)-\(0.281\); Spearman \(\rho=0.222\)-\(0.270\)), indicating limited overlap among the information captured by the three components.
Correlations among the 13 system-level means are higher (Pearson \(r=0.767\)-\(0.879\); Spearman \(\rho=0.687\)-\(0.896\)), showing greater agreement after aggregation at system-level. 
%
%
Together, these findings support retaining the individual alignment components alongside the aggregate MCAS.
\begin{table}[htb]
\caption{Musical-content alignment case study results.}
\label{tab:components}
\vspace{4pt}

\footnotesize
\renewcommand{\arraystretch}{1.05}
\noindent

\begin{minipage}{\columnwidth}
\centering
\textbf{(a) Open systems}
\par\vspace{2pt}

\setlength{\tabcolsep}{2pt}
\begin{tabular*}{\columnwidth}{@{\extracolsep{\fill}}
  >{\raggedright\arraybackslash}p{3.0cm}
  c
  c
  c
  c
@{}}
\hline
\textbf{System} &
\textbf{SA} &
\textbf{GA} &
\textbf{DA} &
\textbf{MCAS} \\
\hline

ACE-Step v1 3.5B
  & 0.337
  & 0.558
  & 0.749
  & 0.517 \\

AudioLDM2 Music
  & 0.229
  & 0.577
  & 0.734
  & 0.477 \\

DiffRhythm
  & 0.358
  & 0.547
  & 0.767
  & 0.526 \\

HeartMuLa OSS 3B
  & 0.290
  & 0.538
  & 0.753
  & 0.493 \\

InspireMusic 1.5B Long
  & 0.239
  & 0.441
  & 0.626
  & 0.406 \\

MusicGen Large
  & 0.288
  & 0.645
  & 0.802
  & 0.541 \\

MusicLDM
  & 0.287
  & 0.583
  & 0.764
  & 0.510 \\

Stable Audio Open 1.0
  & 0.365
  & 0.536
  & 0.823
  & 0.539 \\

YuE
  & 0.110
  & 0.466
  & 0.683
  & 0.378 \\

\hline
\end{tabular*}
\end{minipage}

\vspace{4pt}

\begin{minipage}{\columnwidth}
\centering
\textbf{(b) Commercial systems}
\par\vspace{2pt}

\setlength{\tabcolsep}{2pt}
\begin{tabular*}{\columnwidth}{@{\extracolsep{\fill}}
  >{\raggedright\arraybackslash}p{3.0cm}
  c
  c
  c
  c
@{}}
\hline
\textbf{System} &
\textbf{SA} &
\textbf{GA} &
\textbf{DA} &
\textbf{MCAS} \\
\hline

Lyria 2 (via fal.ai)
  & 0.380
  & 0.647
  & 0.827
  & 0.585 \\

Lyria 3 (via fal.ai)
  & 0.460
  & 0.665
  & 0.868
  & 0.634 \\

Stable Audio 2.5
  & 0.476
  & 0.676
  & 0.829
  & 0.634 \\

Stable Audio 3
  & 0.476
  & 0.662
  & 0.834
  & 0.631 \\

\hline
\end{tabular*}
\end{minipage}

\end{table}

Table~\ref{tab:computational} summarizes computational efficiency across the evaluated systems. Among open-source models, ACE-Step has the lowest latency, RTF, and energy consumption, while MusicGen Large uses the least peak RAM and MusicLDM the least peak VRAM. Thus, no single system performs best on all efficiency measures.
%
Among commercial APIs, Stable Audio~2.5 has the lowest observed latency and RTF, whereas Lyria~3 has the lowest API cost. This shows that faster generation does not necessarily correspond to lower cost.
\begin{table}[htb]
\caption{Computational efficiency case study results.
Latency is reported as mean \(\pm\) SD in seconds, RAM/VRAM in GB,
GPU utilization in \%, GPU energy in Wh, and Cost in USD.}
\label{tab:computational}
\vspace{4pt}

\footnotesize
\renewcommand{\arraystretch}{1.08}
\noindent

\begin{minipage}{\columnwidth}
\centering
\textbf{(a) Open systems}
\par\vspace{2pt}

\setlength{\tabcolsep}{1.25pt}
\begin{tabular*}{\columnwidth}{@{\extracolsep{\fill}}
  l
  r@{$\,\pm\,$}l
  c
  c
  c
  c
  c
@{}}
\hline
\textbf{System} &
\multicolumn{2}{c}{\textbf{Latency}} &
\textbf{RTF}$^*$ &
\textbf{RAM} &
\textbf{VRAM} &
\makecell{\textbf{GPU}\\\textbf{util.}} &
\makecell{\textbf{GPU}\\\textbf{energy}} \\
\hline

ACE-Step v1 3.5B
  & 2.3 & \,\,\,0.03
  & \,\,\,0.1
  & \,\,\,3.2
  & \,\,\,8.8
  & 79.2
  & \,\,\,0.2 \\

AudioLDM2 Music
  & 17.5 & \,\,\,0.05
  & \,\,\,0.6
  & \,\,\,4.1
  & \,\,\,4.5
  & 93.0
  & \,\,\,2.2 \\

DiffRhythm
  & 11.0 & \,\,\,0.07
  & \,\,\,0.4
  & \,\,\,5.2
  & \,\,\,7.0
  & 98.7
  & \,\,\,1.4 \\

HeartMuLa OSS 3B
  & 17.4 & \,\,\,0.08
  & \,\,\,0.6
  & \,\,\,3.1
  & 21.5
  & 88.5
  & \,\,\,1.2 \\

InspireMusic 1.5B Long
  & 20.0 & \,\,\,0.03
  & \,\,\,0.7
  & \,\,\,7.0
  & 11.5
  & 85.0
  & \,\,\,1.2 \\

MusicGen Large
  & 23.7 & \,\,\,0.08
  & \,\,\,0.8
  & 2.5
  & 19.1
  & 92.9
  & \,\,\,1.7 \\

MusicLDM
  & 9.3 & \,\,\,0.01
  & \,\,\,0.3
  & \,\,\,3.1
  & \,\,\,3.2
  & 92.4
  & \,\,\,1.1 \\

Stable Audio Open 1.0
  & 62.0 & \,\,\,0.12
  & \,\,\,2.1
  & \,\,\,3.1
  & 11.0
  & 97.6
  & \,\,\,7.7 \\

YuE
  & 439.5 & 65.86
  & 14.7
  & 19.4
  & 15.6
  & 83.4
  & 40.9 \\

\hline
\end{tabular*}
\end{minipage}

\vspace{2pt}

\parbox{\columnwidth}{%
\footnotesize
\raggedright
$^*$ RTF values report means only; SDs are \(<0.01\), except for YuE (\(2.20\)).
}

\vspace{4pt}

\begin{minipage}{\columnwidth}
\centering
\textbf{(b) Commercial systems}
\par\vspace{2pt}

\setlength{\tabcolsep}{1.25pt}

\begin{tabular*}{\columnwidth}{@{}
  l
  @{\hspace{30pt}}
  r@{$\,\pm\,$}l
  @{\hspace{30pt}}
  r@{$\,\pm\,$}l
  @{\hspace{30pt}}
  c
  @{}}
\hline
\textbf{System} &
\multicolumn{2}{c}{\textbf{Latency}\hspace*{30pt}} &
\multicolumn{2}{c}{\textbf{RTF}\hspace*{30pt}} &
\textbf{Cost} \\
\hline

Lyria 2 (via fal.ai)
  & 36.4 & 11.9
  & 1.1 & 0.36
  & 0.10 \\

Lyria 3 (via fal.ai)
  & 11.6 & \,\,\,1.8
  & 0.4 & 0.06
  & 0.04 \\

Stable Audio 2.5
  & 5.6 & \,\,\,0.6
  & 0.2 & 0.02
  & 0.20 \\

Stable Audio 3
  & 9.8 & \,\,\,2.1
  & 0.3 & 0.07
  & 0.26 \\

\hline
\end{tabular*}
\end{minipage}

\end{table}

\section{Conclusion}
\label{sec:conclusion}
In this paper, we present TTM-Bench, a reference-based framework for benchmarking TTM system performance across musical-content alignment and computational efficiency.
The case study demonstrates the framework's application to 13 TTM systems using a corpus of 100 reference music tracks. The results show that alignment components capture distinct information at the generation level, supporting their individual reporting alongside the aggregate MCAS.
Relative system-level MCAS results remain stable across the tested weighting alternatives, whereas small differences between similarly scoring systems are sensitive to reference-set composition.
Computational results reveal that higher musical-content alignment does not systematically coincide with lower computational demands.
These findings are limited to the evaluated references, system versions, configurations, and access conditions. Moreover, MCAS remains an experimentally defined aggregate whose weighting should be further validated against human judgments.
Future work will extend TTM-Bench to broader reference collections and investigate the relationship between its measures and human judgments of TTM system performance.

\section{Acknowledgment}
No funding was received for conducting this study. The authors have no relevant financial or nonfinancial interests to disclose.

\section{Compliance with Ethical Standards}
This study did not involve human participants or animals; therefore, ethical approval was not required.


\begingroup
\sloppy
\bibliographystyle{IEEEbib}
\bibliography{refs}
\endgroup

\end{document}